\documentclass[aps,prl,reprint,superscriptaddress,amsmath,amssymb,floatfix]{revtex4-2}

\usepackage{graphicx}
\usepackage{dcolumn}
\usepackage{bm}
\usepackage{multirow}
\usepackage{hyperref}
\usepackage[capitalise,noabbrev]{cleveref}
\usepackage{siunitx}
\usepackage{tikz}
\usepackage{physics}
\AtBeginDocument{}

\newcommand{\supp}{\operatorname{supp}}

\begin{document}

\title{Symmetry-Protected Basin Localization in Variational Quantum Eigensolvers}

\author{Yangshuai Wang}
\affiliation{Department of Mathematics, National University of Singapore,
10 Lower Kent Ridge Road, Singapore.}

\begin{abstract}
Variational quantum eigensolvers fail before optimization begins when strong correlation
splits the molecular energy landscape into competing basins and the initial state selects a
non-ground-state basin.
We introduce a geometry-conditioned preconditioner
$\mathcal{P}_{\mathrm{eq}}:\mathbf{R}\mapsto\boldsymbol{\theta}_0$ constrained by the
$SE(3)$ covariance of the molecular Hamiltonian, so that nuclear geometry is mapped directly
into circuit parameters in the correlated ground-state basin.
This basin localization changes the relevant gradient statistics from concentration controlled
to curvature controlled.
In statevector benchmarks on six stretched molecules, $\mathcal{P}_{\mathrm{eq}}$ reduces
Hartree--Fock initialization errors by factors of $38\times$--$6250\times$, reaches sub-mHa
initialization in CO, LiH, and H$_8$, and places N$_2$, H$_2$O, and BeH$_2$ in the mHa-scale
correlated basin.
In disordered H$_{10}$ chains, equivariant basin targeting and stochastic escape reach unit
success probability at fixed optimization budget.
The procedure performs basin selection before the shot-limited quantum loop; the quantum circuit
then refines correlation inside the selected basin.
\end{abstract}

\maketitle

\section{Introduction}
\label{sec:introduction}

\begin{figure}[tb]
  \centering
  \includegraphics[width=0.9\linewidth]{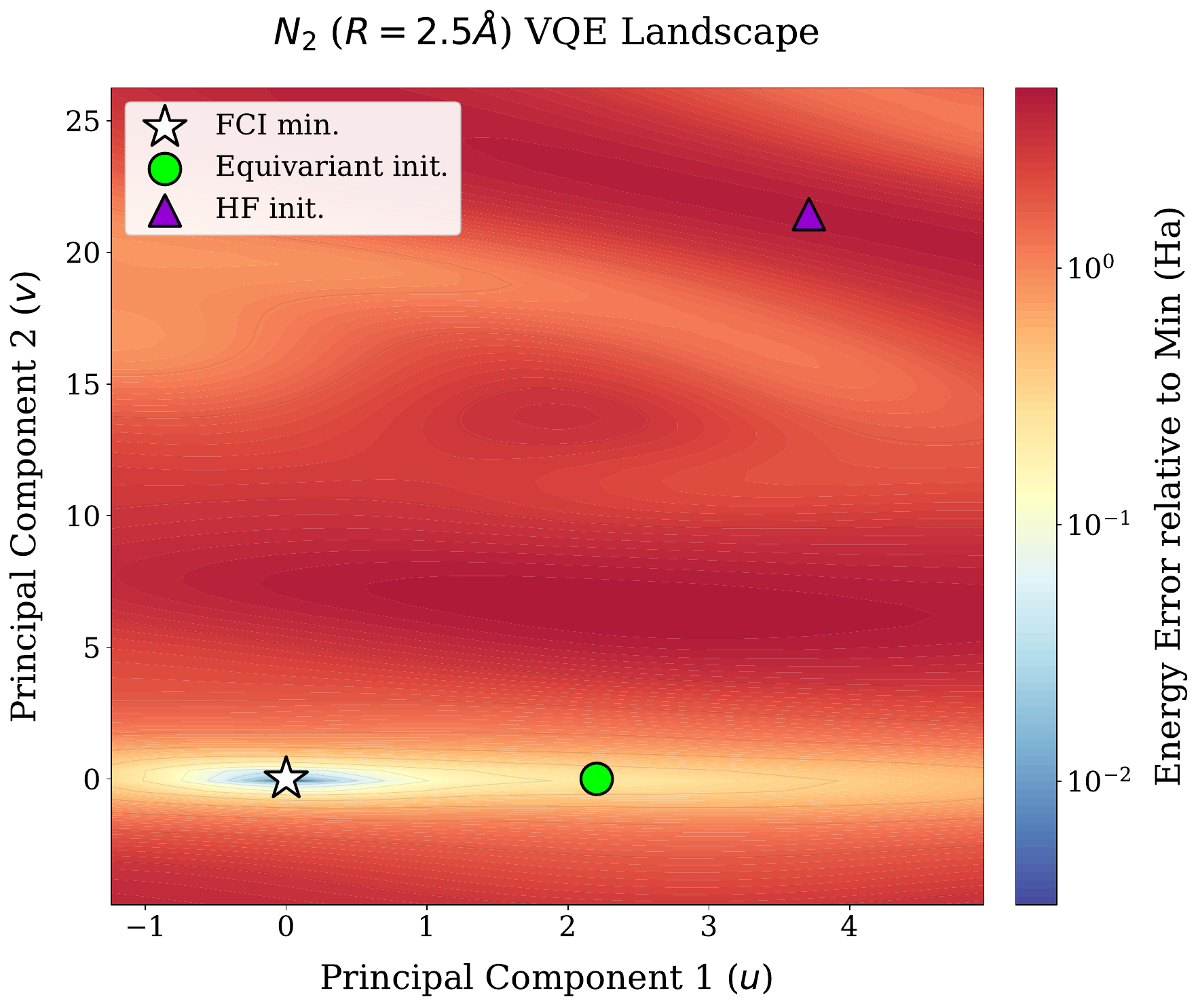}
  \caption{\textbf{Basin topology controls VQE trajectories in the strongly correlated regime.}
  Two-parameter energy landscape of stretched $\mathrm{N}_2$ at $R=2.5$~\AA\ with
  Hartree--Fock and equivariant initialization trajectories overlaid.
  Hartree--Fock lies on the symmetry-broken side of the Coulson--Fischer instability; spin-symmetry breaking diverts
  the variational flow into a competing local minimum ($\Delta E>0.2$~Ha), while
  $\mathcal{P}_{\mathrm{eq}}$ places $\boldsymbol{\theta}_0$ in a locally positive-curvature
  region associated with the correlated ground-state basin (SM~Note~S2).}
  \label{fig:landscape}
\end{figure}

The main obstruction to variational quantum eigensolvers (VQEs) in strongly correlated
molecules is not optimizer inefficiency alone, but basin selection in a nonconvex energy
landscape~\cite{peruzzo2014variational,mcclean2016theory,kandala2017hardware,
cao2019quantum,tilly2022variational,wang2019accelerated,grimsley2019adaptive,
cerezo2021variational,bittel2021training,yang2017optimizing,wang2021noise}.
In strongly correlated regimes~\cite{song2017strongly,dagotto2005complexity}, molecular
symmetry partitions this landscape into competing basins.
The required early step is therefore \emph{basin identification}: placement of
$\boldsymbol{\theta}_0$ inside the ground-state basin of attraction before local optimization
begins (Fig.~\ref{fig:landscape}).

Two distinct mechanisms displace the initialization measure from the locally convex neighborhood
required for curvature-controlled descent.
Random initialization approximates a Haar-like circuit ensemble, where concentration of measure
enforces $\mathrm{Var}(\partial_{\theta_k}E)=O(1/2^N)$, producing exponentially vanishing gradients
that make local descent uninformative~\cite{mcclean2018barren, cerezo2021variational,
larocca2025barren}.
Mean-field initialization fails for the complementary reason: beyond the Coulson--Fischer
point~\cite{coulson1949xxxiv, lyakh2012multireference}, Hartree--Fock spontaneously breaks spin
symmetry, biasing the variational state toward symmetry-broken manifolds with suppressed overlap
with the correlated ground-state sector.
In both cases the initialization measure is displaced away from the basin-attractive neighborhood,
and local descent is diverted into competing local minima; basin identification fails before a single
gradient step is taken.

We show that imposing $SE(3)$ equivariance on the initialization map localizes the induced
measure inside a symmetry-consistent component of the ground-state basin, replacing delocalized
basin discovery by curvature-controlled local descent.
The formal requirement that $\mathcal{P}_{\mathrm{eq}}:\{\mathbf{r}_i\}_{i=1}^M\mapsto\boldsymbol{\theta}_0$
commute with the physical symmetry group of the Hamiltonian restricts the induced initialization
measure to the quotient $\mathbb{R}^{3M}/SE(3)$ and removes redundant rotational and translational
directions from the supervised basin map.

We test the mechanism through four diagnostics.
Fig.~\ref{fig:landscape} resolves the landscape topology: equivariant preconditioning places
$\boldsymbol{\theta}_0$ inside the ground-state basin while Hartree--Fock is diverted into a
symmetry-broken local minimum.
Fig.~\ref{fig:accuracy} and Table~\ref{tab:benchmark} show that this basin placement yields
initialization-error reductions of $38\times$--$6250\times$ across stretched molecular
benchmarks, with sub-mHa accuracy in three transfer systems and mHa-scale localization across
the full benchmark set.
Fig.~\ref{fig:polyatomic} tests transfer to the polyatomic $\mathrm{H_2O}$ O--H stretch,
where the equivariant initialization remains at the mHa scale across geometries.
Fig.~\ref{fig:robustness} shows that under positional disorder equivariance suppresses
large-error tail events, while a hybrid protocol combines deterministic basin targeting with
stochastic escape from residual traps.

\section{Theoretical framework}

We consider electronic-structure Hamiltonians $\hat{H}(\mathbf{R})$ parameterized by nuclear
geometry $\mathbf{R}=\{\mathbf{r}_i\}_{i=1}^{M}\in\mathbb{R}^{3M}$, where $M$ is the number
of nuclei.
In VQE one prepares $\ket{\psi(\boldsymbol{\theta})}$ with a parameterized circuit and minimizes
\begin{equation}
E(\boldsymbol{\theta};\mathbf{R})
  =\bra{\psi(\boldsymbol{\theta})}\hat{H}(\mathbf{R})\ket{\psi(\boldsymbol{\theta})},
\end{equation}
where hats denote operators and ${}^\dagger$ Hermitian conjugation.
Under $g\in SE(3)$ the Hamiltonian transforms covariantly,
\begin{equation}
\hat{U}(g)\,\hat{H}(\mathbf{R})\,\hat{U}^\dagger(g)=\hat{H}(g\cdot\mathbf{R}).
\end{equation}
The initialization $\boldsymbol{\theta}_0$ sets the starting point of this optimization; in
strongly correlated regimes performance is strongly influenced by whether
$\boldsymbol{\theta}_0$ lies inside the ground-state quadratic basin of attraction.

\paragraph{Equivariant preconditioning as a holonomic constraint.}
We define the preconditioner as a structure-preserving map
$\mathcal{P}_{\mathrm{eq}}:\mathbf{R}\mapsto\boldsymbol{\theta}_0\in\Theta_{\mathrm{circuit}}$,
where $\Theta_{\mathrm{circuit}}$ denotes the parameter domain modulo standard ansatz gauge
redundancies (angle periodicities, global-phase gauge).
Because the Hamiltonian is $SE(3)$-covariant, a geometry-conditioned preconditioner is constructed to satisfy
\begin{equation}
\ket{\psi\!\left(\mathcal{P}_{\mathrm{eq}}(g\cdot\mathbf{R})\right)}
\simeq
\hat{U}(g)\ket{\psi\!\left(\mathcal{P}_{\mathrm{eq}}(\mathbf{R})\right)},
\qquad\forall g\in SE(3),
\label{eq:commutativity}
\end{equation}
where $\simeq$ denotes equality of physical states in projective Hilbert space,
modulo ansatz gauge redundancies.
Equation~\eqref{eq:commutativity} imposes a holonomic restriction on admissible
initializations: it removes redundant rigid-body degrees of freedom. The supervised basin target
then fixes the remaining quotient coordinates to a basin-attractive representative in the
symmetry-compatible sector.

\paragraph{Realization.}
$\mathcal{P}_{\mathrm{eq}}$ is implemented as an $SE(3)$-equivariant map from $\mathbf{R}$ to a
basin-attractive parameter representative $\boldsymbol{\theta}^\star(\mathbf{R})$ modulo gauge
redundancies, using a MACE-type equivariant message-passing network built from the same
tensor-coupling principles used in $E(3)$-equivariant atomistic models~\cite{thomas2018tensor,
batzner2022nequip,schutt2021equivariant,fuchs2020se3,geiger2022e3nn,batatia2022mace,
wang2025acegraphene}.
A finite-range equivariant map and a depth-$L$ locality-preserving ansatz share the same
receptive field set by the circuit causal cone, so locality constraints governing ansatz
expressibility are matched by construction (SM~Notes~S3--S4).
Numerically, localized-orbital Hamiltonians are built with PySCF~\cite{sun2018pyscf};
offline statevector target labels are obtained by basin-hopping followed by L-BFGS-B
refinement~\cite{byrd1995limited}; and the network is trained on periodic angle
representatives with held-out geometry tests and few-label readout adaptation, following
machine-learned interatomic-potential generalization and fine-tuning practice~\cite{ortner2023framework,wang2024theoretical,
liu2026finetuning} (SM~Note~S7).

\paragraph{From concentration-controlled to curvature-controlled gradients.}
Standard barren-plateau arguments assume random initialization realizes a unitary $t$-design on
$d=2^N$ dimensions, so concentration of measure gives
$\mathrm{Var}(\partial_{\theta_k}E)=O(d^{-1})$~\cite{mcclean2018barren, cerezo2021variational}.
Basin localization changes the measure to which the concentration estimate applies.
Physically, the two regimes correspond to: (i)~design-like initialization, where the circuit
ensemble explores the full unitary volume and measure concentration suppresses gradients
exponentially in system size; and (ii)~basin-localized initialization, where the measure's
support is confined to a strongly convex neighborhood (in non-redundant coordinates) and
gradients reflect local Hessian curvature rather than Hilbert-space volume.

\medskip
\noindent\textbf{Proposition~1} (Curvature-controlled vs.\ concentration-controlled gradients).
\textit{Let $\mu$ denote the initialization measure and $d_\lambda$ the Hilbert dimension of the
relevant symmetry sector.
If $\mu$ is design-like on that sector, concentration of measure implies}
\begin{equation}
\mathrm{Var}_{\mu}\!\left(\partial_{\theta_k}E\right)=O(\mathrm{poly}(N)/d_\lambda).
\end{equation}
\textit{If instead $\supp(\mu)\subset B$, where $B$ is a strongly convex quadratic basin with
$\kappa_m\mathbf{I}\preceq\mathcal{H}(\boldsymbol{\theta})\preceq \kappa_M\mathbf{I}$ for all
$\boldsymbol{\theta}\in B$ in non-redundant coordinates, then gradients are curvature-controlled.
For any non-redundant unit direction $u$,}
\begin{equation}
\kappa_m^2\,\lambda_{\min}(\Sigma)
\;\lesssim\;
\mathrm{Var}_{\mu}\!\left(u^\top\nabla E\right)
\;\lesssim\;
\kappa_M^2\,\lambda_{\max}(\Sigma),
\end{equation}
\textit{where $\Sigma$ is the covariance matrix of $\mu$ in basin coordinates, restricted to the
non-redundant support.
Equivalently, in local Hessian eigenmodes,}
\begin{equation}
\mathrm{Var}_{\mu}(g_j)\simeq \kappa_j^2\sigma_j ,
\end{equation}
\textit{with $\kappa_j$ the local curvature and $\sigma_j$ the initialization variance along mode
$j$.
A symmetry sector can still concentrate under a design-like measure; localization inside a
quadratic basin changes the scaling.}
The gradient variance thus changes from exponentially suppressed ($\sim d_\lambda^{-1}$,
design-like) to Hessian-mode variance $\kappa_j^2\sigma_j$ (basin-localized).

\noindent\emph{Proof: SM~Note~S1.}\medskip

\section{Results}

\subsection*{a.~SE(3)-equivariance resolves basin identification before the first quantum evaluation}

Figure~\ref{fig:landscape} makes the basin-topology obstruction explicit for stretched
$\mathrm{N}_2$ at $R=2.5$~\AA, a canonical strong-correlation regime where the Coulson--Fischer
instability partitions the landscape into topologically disconnected basins.
The Hartree--Fock seed lies on the symmetry-broken side of this instability; spin-symmetry breaking diverts the variational
flow into a competing local minimum ($\Delta E>0.2$~Ha) with strongly suppressed overlap with the
correlated singlet sector.
The basin boundary is a symmetry-enforced obstruction that local descent from that initial
condition does not cross.

By contrast, $\mathcal{P}_{\mathrm{eq}}$ places $\boldsymbol{\theta}_0$ inside the
ground-state quadratic basin, where the Hessian is non-negative up to near-zero modes attributable
to gauge redundancies.
Basin membership is assessed through the local quadratic expansion
\begin{equation}
E(\boldsymbol{\theta}_0+\delta\boldsymbol{\theta})
  \approx E(\boldsymbol{\theta}_0)
  +\nabla E(\boldsymbol{\theta}_0)^\top\delta\boldsymbol{\theta}
  +\tfrac{1}{2}\delta\boldsymbol{\theta}^\top\mathcal{H}(\boldsymbol{\theta}_0)\,
   \delta\boldsymbol{\theta},
\label{eq:quadratic}
\end{equation}
with the Hessian eigenspectrum as a curvature diagnostic.
Standard initialization exhibits negative-curvature eigenvalues, indicating escape directions
toward competing basins.
Equivariant initialization instead produces a spectrum that is positive apart from near-zero gauge
modes; after projection to non-redundant coordinates this identifies a strongly convex
neighborhood of the correlated minimum
(SM~Note~S2).

\subsection*{b.~Chemical-accuracy initialization from symmetry-protected basin placement}

\begin{figure*}[t]
\centering
\includegraphics[width=\linewidth]{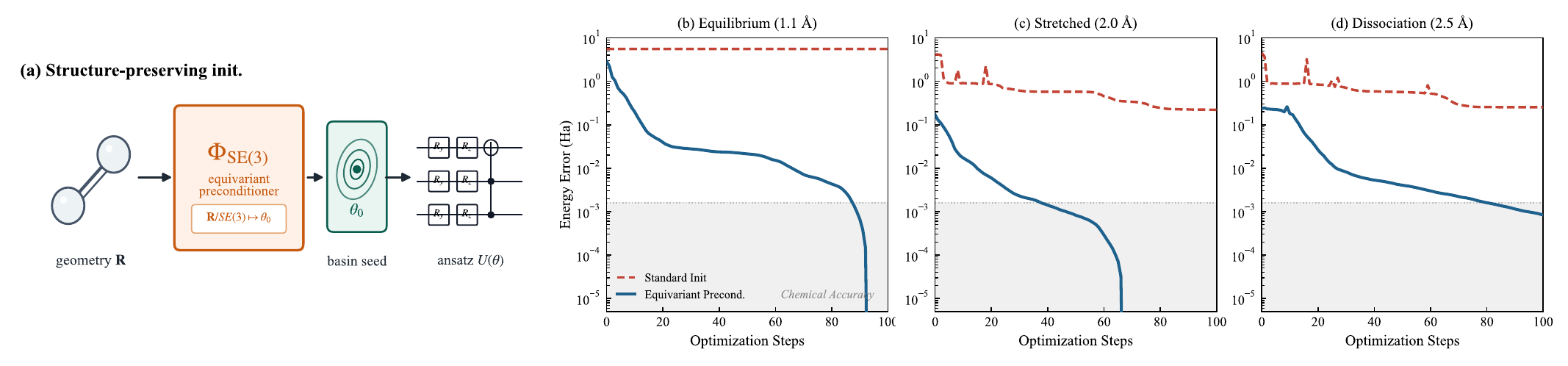}
\caption{\label{fig:accuracy}
\textbf{Equivariant preconditioning places circuit parameters on the correlated branch of
the PES at initialization, before quantum optimization.}
(a)~$\mathcal{P}_{\mathrm{eq}}$ maps geometry $\mathbf{R}$ to symmetry-consistent
$\boldsymbol{\theta}_{\mathrm{init}}$ via a single classical evaluation;
(b--d)~energy-error traces during variational refinement of $\mathrm{N}_2$ at equilibrium,
stretched, and dissociation geometries, comparing Hartree--Fock and equivariant initialization
against the statevector reference.
When the optimizer terminates before 100 recorded calls, the final evaluated error is continued
horizontally to show the terminal plateau.
In stretched regimes, Hartree--Fock enters a symmetry-broken basin
($\Delta E\gtrsim0.2$~Ha); equivariant preconditioning lowers the initial error to the
chemical-accuracy threshold in the most transferable cases and to the mHa-scale correlated
basin across the benchmark set. Subsequent quantum refinement is then a
within-basin optimization (SM~Note~S2).}
\end{figure*}

Figure~\ref{fig:accuracy} shows that basin placement translates directly into energetic
accuracy: across the correlated branch of the PES for stretched $\mathrm{N}_2$ and
$\mathrm{H_2O}$, $\mathcal{P}_{\mathrm{eq}}$ tracks the reference branch at initialization
while Hartree--Fock initialization incurs errors of order $10^{-1}$--$1$~Ha by entering the
symmetry-broken basin.
The basin assignment is fixed before any shot-based evaluation: one classical evaluation of
$\mathcal{P}_{\mathrm{eq}}$ sets the initialization from the multi-reference basin structure
implied by molecular geometry and symmetry.

Table~\ref{tab:benchmark} quantifies the improvement across six strongly correlated molecular
benchmarks.
The factors of $38\times$--$6250\times$ measure a change from symmetry-broken-basin placement to
correlated-basin placement.
$\mathcal{P}_{\mathrm{eq}}$ encodes the coarse-grained multi-reference structure that is visible
from geometry and symmetry; the quantum processor is then used for within-basin refinement of
residual entanglement.

\begin{table}
\caption{\label{tab:benchmark}
\textbf{Transfer of basin-localized initialization across stretched molecular benchmarks.}
Initialization energy errors relative to the reference ground-state energy,
$\Delta E=E_{\mathrm{init}}-E_{\mathrm{ref}}$.
In stretched geometries Hartree--Fock incurs large errors due to symmetry breaking and absent
multi-reference structure; equivariant preconditioning reduces $\Delta E$ by
$38\times$--$6250\times$. Bold rows mark systems where $\mathcal{P}_{\mathrm{eq}}$ achieves
sub-mHa initialization accuracy.}
\begin{ruledtabular}
\begin{tabular}{lcccc}
System & Geometry & $\Delta E_{\text{HF}}$ (Ha)
       & $\Delta E_{\mathcal{P}_{\mathrm{eq}}}$ (Ha) & Improv. \\
N$_2$          & 2.5~\AA & $4.479$ & $0.0074$          & $604\times$          \\
H$_2$O         & 2.5~\AA & $0.613$ & $0.0162$          & $38\times$          \\
BeH$_2$        & 2.5~\AA & $0.781$ & $0.0059$          & $132\times$          \\
\textbf{CO}    & \textbf{2.5~\AA} & $0.257$
               & $\mathbf{0.00068}$ & $\mathbf{378\times}$ \\
\textbf{LiH}   & \textbf{3.0~\AA} & $0.463$
               & $\mathbf{0.00050}$ & $\mathbf{924\times}$  \\
\textbf{H$_8$} & \textbf{2.5~\AA} & $2.425$
               & $\mathbf{0.00039}$ & $\mathbf{6250\times}$ \\
\end{tabular}
\end{ruledtabular}
\end{table}

\subsection*{c.~Polyatomic transfer beyond hydrogen chains}

\begin{figure}[tb]
\centering
\includegraphics[width=0.9\linewidth]{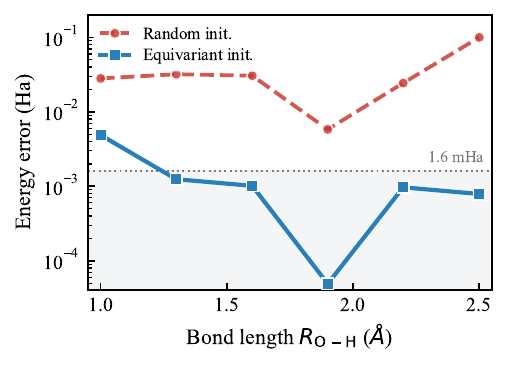}
\caption{\label{fig:polyatomic}
\textbf{Polyatomic transfer under equivariant preconditioning.}
Energy error for the $\mathrm{H_2O}$ O--H stretch after geometry-conditioned initialization;
the shaded region and dotted line mark chemical accuracy.
Colors compare random initialization (red) with equivariant initialization
(blue).}
\end{figure}

Figure~\ref{fig:polyatomic} tests the same initialization map away from linear hydrogen chains.
For the $\mathrm{H_2O}$ O--H stretch, equivariant preconditioning keeps the initial state at the
mHa scale and inside chemical accuracy for most stretched geometries, while random initialization
remains one to two orders of magnitude larger across the scan.

\subsection*{d.~Hybrid basin targeting improves success under geometric disorder}

\begin{figure}[tb]
\centering
\includegraphics[width=0.9\linewidth]{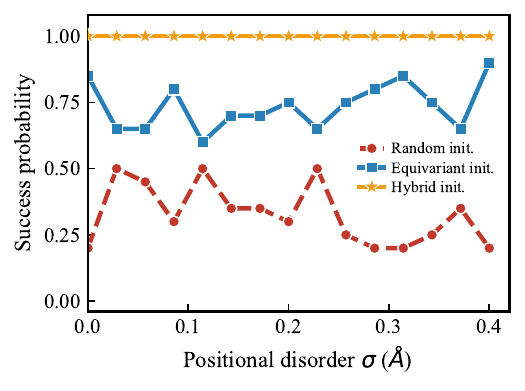}
\caption{\label{fig:robustness}
\textbf{Hybridizing geometric basin targeting with stochastic escape improves ground-state
success under positional disorder.}
Ground-state success probability for $\mathrm{H}_{10}$ chains vs.\ positional disorder
strength $\sigma$ at fixed optimization budget, for random (red), equivariant (blue),
and hybrid (gold) initialization strategies.
Equivariant preconditioning suppresses heavy-tail basin misidentification failures while
stochasticity resolves fine-scale inter-well traps; their combination reaches success probability
one across the disorder scan (SM~Figs.~S1--S2).}
\end{figure}

Figure~\ref{fig:robustness} shows that positional disorder fragments the variational
landscape and changes basin identification quality.
Random initialization degrades with disorder strength $\sigma$; the associated error statistics
become heavy-tailed, with rare large-error basin misidentification events dominating the
failure rate.
Equivariant preconditioning suppresses the far tail; in these tests,
local geometric information targets the ground-state basin even in the absence
of global translational symmetry.

The hybrid method reaches success probability one across the disorder scan, indicating a
complementarity between equivariant basin targeting and stochastic escape.
Equivariance suppresses the heavy-tail failure mode by concentrating the initialization measure
onto symmetry-consistent, basin-attractive coordinates; stochasticity then resolves the
fine-scale disorder-induced traps within that basin-localized sector (SM~Figs.~S1--S2).

\section{Discussion}

\paragraph{Classical initialization cost.}
At fixed target precision $\epsilon$, shot-based estimation contributes $O(\epsilon^{-2})$
cost per observable, while classical evaluation of $\mathcal{P}_{\mathrm{eq}}$ requires
$O(1)$ milliseconds, a wall-clock ratio of $10^4$--$10^6$ in favor of the classical
initialization step.
In the strong-correlation benchmarks studied here, the preconditioner is small relative to the
quantum measurement overhead: basin identification is shifted from the shot-limited quantum loop
to a symmetry-protected classical map.

\paragraph{Generality: symmetry class, Hamiltonians, and ans\"atze.}
The construction is defined for electronic-structure Hamiltonians whose symmetry group contains
$SE(3)$ as a subgroup, including non-relativistic molecular Hamiltonians
in the Born--Oppenheimer approximation.
The holonomic constraint~\eqref{eq:commutativity} is ansatz-agnostic: it constrains the
initialization measure independently of whether the circuit is hardware-efficient, UCCSD-type,
or a locally connected shallow ansatz.
The locality matching between the equivariant receptive field and the circuit causal cone
is specific to depth-$L$ locality-preserving ans\"atze; deeper globally entangling
ans\"atze require receptive fields commensurate with their causal cones and therefore define a
different geometric preconditioning problem (SM~Note~S3).
Extension to periodic systems with space-group symmetry and to spin-orbit-coupled Hamiltonians
with $SU(2)$-breaking requires equivariant maps compatible with the enlarged symmetry group.

\paragraph{Static versus dynamic correlation.}
The results separate two sources of correlation.
$\mathcal{P}_{\mathrm{eq}}$ encodes the coarse-grained multi-reference basin structure dictated by
molecular geometry and symmetry, a form of effective static-correlation inference that is cheap
classically but costly to discover through shot-based search.
Quantitatively, $\mathcal{P}_{\rm eq}$ resolves the multi-reference basin structure with
sub-mHa initialization in CO, LiH, and H$_8$ and mHa-scale localization across all six stretched
benchmarks (Table~\ref{tab:benchmark}), while leaving the quantum processor for
within-basin dynamical refinement (Fig.~\ref{fig:accuracy}b--d).
The shot-limited loop is not used for geometric search over disconnected basins; it is reserved
for controlled corrections within the physically relevant symmetry sector.

\paragraph{Barren plateaus as a measure-support mechanism.}
This view recasts barren plateaus as a property of \emph{delocalized} initialization measures
whose induced ensembles acquire design-like statistics, rather than as an unavoidable consequence
of many-body Hilbert-space dimension.
$SE(3)$-equivariance restricts the induced ensemble to symmetry-consistent, basin-localized
support where gradient statistics are set by local Hessian geometry.
The asymptotic separation is the content of Proposition~1:
design-like support gives
$\mathrm{Var}(\partial_{\theta_k}E)=O(\mathrm{poly}(N)/d_\lambda)$, with $d_\lambda=2^N$ on the full
Hilbert space, whereas basin-localized support gives a curvature-controlled bound independent of
the full Hilbert volume.
\paragraph{Depth and transfer.}
At a fixed working depth, the operational requirement is that the initialization measure retain
basin-localized support rather than spreading into design-like statistics.
For larger devices, the relevant threshold depth $L^\star(N)$ is the point at which the induced
initialization measure loses basin-localized support and begins to acquire design-like statistics.
$\mathcal{P}_{\mathrm{eq}}$ is trained on a finite geometry set via a MACE-type equivariant
network; the map encodes basin correspondence rather than energy interpolation, and the
few-label adaptation protocol uses
$N_{\mathrm{adapt}}\approx10$--$50$ labeled geometries in the transfer protocols reported in
SM~Note~S5.
In the disorder tests, equivariance supplies deterministic basin targeting, while stochastic
restarts provide escape from residual fine-scale traps.

\paragraph{Outlook.}
Equivariant preconditioning defines a geometry-conditioned VQE protocol:
the classical map selects the symmetry-compatible basin, and the quantum loop performs
basin-local energy minimization.
The same principle specifies enlarged physical symmetry groups: periodic systems require
equivariant maps compatible with space-group symmetry, where the relevant equivalence class is
$\mathbb{R}^{3M}/(\text{space group})$ rather than $\mathbb{R}^{3M}/SE(3)$; spin-orbit-coupled
Hamiltonians with $SU(2)$-breaking require maps that transform covariantly under the double
cover, imposing a distinct equivariance constraint on the initialization measure.

\clearpage
\onecolumngrid
\begin{center}
{\large\bf Supplemental Material for ``Symmetry-Protected Basin Localization in Variational Quantum Eigensolvers''}
\end{center}

\noindent This Supplemental Material is organized as follows.
Note~S1 derives Proposition~1 and delineates the design-like versus basin-localized
initialization regimes underlying the gradient-variance discussion.
Note~S2 reports Hessian eigenspectra and near-zero redundant modes as curvature diagnostics of
basin membership.
Note~S3 details the $SE(3)$-equivariant realization of $\mathcal{P}_{\mathrm{eq}}$
(MACE parameterization and equivariance verification).
Note~S4 provides the variational ansatz and geometry-to-circuit parameterization, including
reservoir structure and symmetry-sector decomposition.
Note~S5 presents ablations isolating the equivariance contribution, data-efficiency benchmarks,
and analysis of performance in the data-sparse and out-of-distribution geometry regimes.
Note~S6 gives resource accounting: shot-cost model, step complexity, and classical overhead
analysis at fixed precision.
Note~S7 summarizes reproducibility details.

\setcounter{equation}{0}
\setcounter{figure}{0}
\setcounter{table}{0}
\setcounter{section}{0}

\renewcommand{\theequation}{S\arabic{equation}}
\renewcommand{\thefigure}{S\arabic{figure}}
\renewcommand{\thetable}{S\arabic{table}}
\renewcommand{\thesection}{Note S\arabic{section}}
\renewcommand{\theHfigure}{S\arabic{figure}}
\renewcommand{\theHtable}{S\arabic{table}}
\renewcommand{\theHequation}{S\arabic{equation}}
\renewcommand{\theHsection}{S\arabic{section}}

\section{Symmetry-protected gradient statistics: derivation of Proposition~1}
\label{note:theory}

This note derives curvature-controlled bounds on the gradient variance and contrasts two regimes:
(i) a Haar-like (unitary-design) ensemble in which concentration of measure enforces
$\mathrm{Var}(\partial_k E)=O(1/d)$, and
(ii) a basin-localized ensemble in which gradient statistics are governed by local curvature
rather than Hilbert-space volume.
The result stated here as Theorem~S1 is the formal basis of Proposition~1 in the main text.

\subsection{Formal setup: gradient operator}

Consider a parameterized ansatz $\ket{\psi(\boldsymbol{\theta})}=U(\boldsymbol{\theta})\ket{\psi_{\mathrm{ref}}}$
with energy functional
$E(\boldsymbol{\theta})=\bra{\psi(\boldsymbol{\theta})}\hat{H}\ket{\psi(\boldsymbol{\theta})}$.
For the $k$-th parameter,
\begin{equation}
\partial_k E(\boldsymbol{\theta})
= \bra{\psi(\boldsymbol{\theta})}\,\hat{G}_k\,\ket{\psi(\boldsymbol{\theta})},
\qquad
\hat{G}_k:=i[\hat{A}_k,\hat{H}],
\label{eq:S1_grad_comm}
\end{equation}
where $\hat{A}_k:=-i\,U^\dagger(\boldsymbol{\theta})\,\partial_{\theta_k}U(\boldsymbol{\theta})$ is the
Hermitian generator associated with $\theta_k$ (for a Pauli-rotation gate
$U=e^{i\theta_k\hat{\sigma}_k/2}$ one has $\hat{A}_k=\hat{\sigma}_k/2$).
The gradient observable $\hat{G}_k$ is Hermitian and traceless. For Pauli-rotation gates,
$\|\hat{A}_k\|_\infty\le 1/2$.

\subsection{Haar-like (unitary-design) regime}

For deep, symmetry-agnostic circuits whose distribution over states approaches a unitary $2$-design on a
Hilbert space of dimension $d$ (with $d=2^N$, or $d=d_\lambda$ if restricted to a symmetry sector),
Weingarten calculus yields (see, e.g.,~\cite{mcclean2018barren} and references therein)
\begin{equation}
\mathrm{Var}_{\mathrm{Haar}}\!\left[\partial_k E\right]
=\frac{\overline{\mathrm{tr}}(\hat{G}_k^2)-
\overline{\mathrm{tr}}(\hat{G}_k)^2}{d+1}
\le
\frac{\overline{\mathrm{tr}}(\hat{G}_k^2)}{d+1},
\label{eq:S1_haar_var}
\end{equation}
where $\overline{\mathrm{tr}}(\cdot)=d^{-1}\mathrm{Tr}(\cdot)$ denotes the normalized trace.
Tracelessness follows from cyclicity,
$\mathrm{Tr}\,\hat{G}_k=i\mathrm{Tr}(\hat{A}_k\hat{H}-\hat{H}\hat{A}_k)=0$.
For local Hamiltonians and local generators,
$\overline{\mathrm{tr}}(\hat{G}_k^2)$ grows at most polynomially in $N$ while $d$ grows
exponentially, so $\mathrm{Var}_{\mathrm{Haar}}[\partial_k E]=O(\mathrm{poly}(N)/d)$: an exponential suppression arising from
the delocalized support of the initialization measure over the unitary manifold.

\subsection{Basin-localized regime: Theorem~S1}

\textbf{Assumption~A (strongly convex basin).}
There exists a neighborhood $\mathcal{B}\subset\mathbb{R}^P$ around a local minimizer
$\boldsymbol{\theta}^\star$ such that
\begin{equation}
\kappa_m\,\mathbf{I} \preceq \nabla^2 E(\boldsymbol{\theta}) \preceq \kappa_M\,\mathbf{I},
\qquad \forall\,\boldsymbol{\theta}\in\mathcal{B},
\label{eq:S1_strong_convex}
\end{equation}
for constants $0<\kappa_m\le \kappa_M<\infty$, interpreted on non-redundant directions when the ansatz
carries gauge degeneracies.

\medskip
\noindent\textbf{Theorem~S1} (Curvature-controlled gradient variance).
\textit{Let $\boldsymbol{\theta}_0$ be drawn from a distribution supported in $\mathcal{B}$, with
mean $\mathbb{E}[\boldsymbol{\theta}_0]=\boldsymbol{\theta}^\star$ and covariance
$\Sigma=\mathbb{E}\!\left[(\boldsymbol{\theta}_0-\boldsymbol{\theta}^\star)
(\boldsymbol{\theta}_0-\boldsymbol{\theta}^\star)^\top\right]$ on the non-redundant subspace.
In the local quadratic basin, for any unit direction $u$ in that subspace,}
\begin{equation}
\kappa_m^2\,\lambda_{\min}(\Sigma)
\ \lesssim\
\mathrm{Var}\!\left[u^\top\nabla E(\boldsymbol{\theta}_0)\right]
\ \lesssim\
\kappa_M^2\,\lambda_{\max}(\Sigma).
\label{eq:S1_var_bounds}
\end{equation}
\textit{Equivalently, in the local Hessian eigenbasis,
$\mathrm{Var}(g_j)\simeq \kappa_j^2\sigma_j$, where $\kappa_j$ is the curvature and
$\sigma_j$ is the initialization variance along eigenmode $j$.}

\textit{Proof.}
Let $\delta\boldsymbol{\theta}:=\boldsymbol{\theta}_0-\boldsymbol{\theta}^\star$.
Inside the quadratic basin,
\begin{equation}
\nabla E(\boldsymbol{\theta}_0)
=H_\star\,\delta\boldsymbol{\theta}+r(\delta\boldsymbol{\theta}),
\qquad
H_\star:=\nabla^2E(\boldsymbol{\theta}^\star),
\end{equation}
with $\|r(\delta\boldsymbol{\theta})\|=O(\|\delta\boldsymbol{\theta}\|^2)$.
The leading gradient covariance is therefore
\begin{equation}
\Gamma_g
=\mathrm{Cov}[\nabla E(\boldsymbol{\theta}_0)]
=H_\star\Sigma H_\star+O(\mathbb{E}\|\delta\boldsymbol{\theta}\|^3).
\end{equation}
For any unit vector $u$,
\begin{equation}
\mathrm{Var}\!\left[u^\top\nabla E(\boldsymbol{\theta}_0)\right]
=u^\top H_\star\Sigma H_\star u+O(\mathbb{E}\|\delta\boldsymbol{\theta}\|^3).
\end{equation}
Since $\kappa_m I\preceq H_\star\preceq\kappa_M I$ on the non-redundant subspace,
$\kappa_m\le\|H_\star u\|\le\kappa_M$.
The Rayleigh bounds for $\Sigma$ give
\begin{equation}
\lambda_{\min}(\Sigma)\|H_\star u\|^2
\le
u^\top H_\star\Sigma H_\star u
\le
\lambda_{\max}(\Sigma)\|H_\star u\|^2,
\end{equation}
which yields Eq.~\eqref{eq:S1_var_bounds} up to the quadratic-basin remainder.
If $u$ is a local Hessian eigenmode with curvature $\kappa_j$ and $\Sigma$ is resolved in the same
basin coordinates, the leading term reduces to
$\mathrm{Var}(g_j)=\kappa_j^2\sigma_j$.\hfill$\square$

\subsection{Corollary: curvature control under basin localization}

\textbf{Corollary~S1.}
\textit{Suppose $\mathcal{P}_{\mathrm{eq}}$ induces an initialization distribution that
(i) remains within a fixed symmetry sector $\mathcal{H}_\lambda$, and
(ii) is supported inside a strongly convex neighborhood $\mathcal{B}$ of the ground-state minimizer
satisfying Assumption~A.
Then initialization gradients are curvature-controlled in the sense of Theorem~S1.
The Haar-like $d^{-1}$ variance scaling does not apply to this basin-supported measure.}

Conditions (i)--(ii) are supported numerically by the Hessian diagnostics of Note~S2 and the landscape
localization shown in the main text.
Together, Theorem~S1 and Corollary~S1 constitute the proof of Proposition~1 stated in the main text.

\subsection{Symmetry restriction vs.\ geometric localization}

Two reductions of the search space must be distinguished.

\textbf{(i) Symmetry restriction alone does not remove concentration.}
Restricting to a symmetry sector $\mathcal{H}_\lambda$ reduces the accessible Hilbert space from
dimension $2^N$ to $d_\lambda$, but $d_\lambda$ remains exponential in $N$ for generic many-body settings.
A random (Haar-like) search \emph{within} the sector can therefore still exhibit barren-plateau
scaling~\cite{larocca2025barren}.

\textbf{(ii) Geometric localization via equivariant preconditioning.}
Equivariant preconditioning induces a basin-localized initialization measure supported on
symmetry-consistent, basin-attractive coordinates of the variational manifold.
In this regime gradient statistics are governed by local curvature through Theorem~S1, rather than by
the exponential Hilbert-space volume.
For the shallow constant-depth circuits studied here, this replaces delocalized initialization by
classical basin identification through $\mathcal{P}_{\mathrm{eq}}$.

\section{Landscape diagnostics: curvature spectra and heavy-tail suppression}
\label{note:topology}

Note~S1 derives curvature-controlled gradient statistics from initialization inside a
strongly convex basin (Assumption~A).
Here we test this premise numerically using two complementary probes: exact Hessian spectra at
initialization, and large-sample error distributions under disorder.

\subsection{Statistical suppression of heavy-tail errors under disorder}

Heavy-tail error statistics indicate landscape fragmentation: rare large-error initializations
into high-energy regions dominate the failure rate.
To quantify whether equivariant preconditioning suppresses such outliers, we analyze the distribution
of initialization errors $\Delta E:=E(\boldsymbol{\theta}_0)-E_{\mathrm{ref}}$ over many independent
disorder instances (Fig.~\ref{fig:S1_diagnostics}(a)).

\paragraph{Sampling protocol.}
We generate $N_{\mathrm{sample}}=5{,}000$ independent instances for $\mathrm{H}_4$ and
$\mathrm{H}_{10}$ chains.
This sample size is chosen to reliably resolve the tail of rare outlier events: for an outlier
probability $p_{\mathrm{out}}\ll 1$, tail estimation requires $N_{\mathrm{sample}}\gg
p_{\mathrm{out}}^{-1}$.
For equivariant preconditioning, input geometries are additionally perturbed by independent
per-atom Gaussian noise $\delta\mathbf{R}_i\sim\mathcal{N}(\mathbf{0},\,\sigma_{\mathrm{pos}}^2\mathbf{I}_3)$
with $\sigma_{\mathrm{pos}}=0.05$~\AA\ to emulate test-set variation.

\paragraph{Observation.}
Random initialization produces a broad distribution with pronounced heavy tails extending to large
$\Delta E$, indicating frequent sampling of high-energy, symmetry-broken landscape regions.
Equivariant preconditioning narrows the distribution, suppresses the far tail, and concentrates
probability mass at chemical-accuracy scale.
This tail suppression shows that $\mathcal{P}_{\mathrm{eq}}$ concentrates the initialization
measure near basin-attractive coordinates, reducing the probability of sampling
symmetry-incompatible high-energy manifolds in the presence of geometric disorder.

\subsection{Exact Hessian spectrum at initialization}

\begin{figure}[tbp]
\centering
\includegraphics[width=0.92\linewidth]{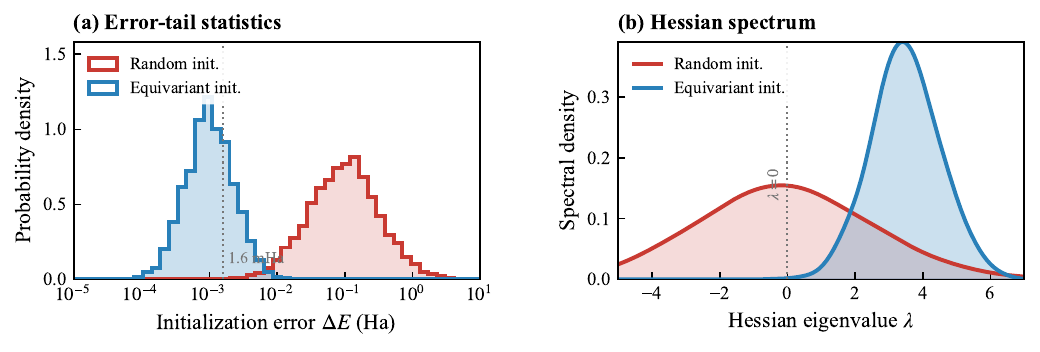}
\caption{\textbf{Landscape diagnostics supporting basin-localized initialization.}
(a)~\textbf{Heavy-tail suppression under equivariant preconditioning.}
Probability density of initialization errors $\Delta E$ for $\mathrm{H}_4$ under positional disorder.
Random initialization (red) exhibits heavy-tailed outliers; equivariant preconditioning (blue)
suppresses rare high-error events, matching basin-localized measure support.
(b)~\textbf{Curvature diagnostic at initialization for stretched $\mathrm{N}_2$.}
Empirical spectral density of the Hessian $\mathcal{H}$ at $\boldsymbol{\theta}_0$.
Random initialization (red) carries negative-curvature weight, indicating saddle-dominated local
geometry; equivariant initialization (blue) yields a non-negative spectrum up to numerical precision.
Near-zero modes reflect ansatz parameter redundancies and gauge directions.}
\label{fig:S1_diagnostics}
\end{figure}

We compute the Hessian $\mathcal{H}\in\mathbb{R}^{P\times P}$ with entries
$\mathcal{H}_{ij}=\partial_{\theta_i}\partial_{\theta_j}E(\boldsymbol{\theta}_0)$ exactly via
automatic differentiation in a statevector simulator, avoiding finite-difference noise
(Fig.~\ref{fig:S1_diagnostics}(b)).

\paragraph{Instance.}
Stretched $\mathrm{N}_2$ at $R=2.5$~\AA, mapped to $12$ qubits with a depth-$L=4$ hardware-efficient
ansatz ($P=120$ parameters).
We diagonalize $\mathcal{H}$ and report the spectral density
$\rho(\lambda)=\frac{1}{P}\sum_{j=1}^{P}\delta(\lambda-\lambda_j)$.

\paragraph{Observation.}
Symmetry-agnostic initialization exhibits substantial negative-curvature weight ($\lambda<0$),
identifying saddle-dominated local geometry and escape directions toward competing basins.
Equivariant initialization ($\mathcal{P}_{\mathrm{eq}}$) yields a spectrum that is non-negative up to
numerical precision; the cluster of near-zero modes reflects parameter redundancies and gauge directions
of the ansatz.
The remaining non-redundant directions carry predominantly positive curvature, supporting Assumption~A.

\section{Equivariant realization of $\mathcal{P}_{\mathrm{eq}}$ (MACE parameterization)}
\label{note:mace}

This note specifies an explicit $SE(3)$-equivariant parameterization of the structure-preserving map
$\mathcal{P}_{\mathrm{eq}}:\mathbf{R}\mapsto\boldsymbol{\theta}_0$.
We use an equivariant message-passing architecture (MACE~\cite{batatia2022mace}) for this
realization, in the lineage of $E(3)$-equivariant tensor message-passing models for atomistic
systems~\cite{thomas2018tensor,batzner2022nequip,schutt2021equivariant,fuchs2020se3,
geiger2022e3nn}. The ingredients used for basin localization are:
(i) exact $SE(3)$ equivariance of intermediate tensor features,
(ii) a local readout matched to the shallow ansatz causal cone (Note~S4), and
(iii) a training objective that respects gauge redundancies of the circuit parameterization.
These choices follow machine-learned interatomic-potential (MLIP) practice in which locality, body
order, and the choice of training observations control transfer outside the labeled
set~\cite{chen2022qmml,ortner2023framework,wang2024theoretical}.

\subsection{Graph construction and atomic embeddings}

The molecular geometry is represented as a neighbor graph with edges $(i,j)$ whenever
$r_{ij}<r_{\mathrm{cut}}=5.0$~\AA.
Each atom $i$ is assigned an initial feature vector via a learnable embedding of its atomic number $Z_i$
into a 128-dimensional channel space ($K_{\mathrm{ch}}=128$),
\begin{equation}
h^{(0)}_{i,k,0,0}=\sum_{z\in\mathcal{Z}} W_{kz}\,\delta_{zZ_i},
\end{equation}
where $\mathcal{Z}$ is the set of atomic numbers present in the dataset,
$W\in\mathbb{R}^{K_{\mathrm{ch}}\times|\mathcal{Z}|}$ is a learnable embedding matrix,
and the subscript indices denote channel $k$ and spherical-harmonic degree $(\ell=0,m=0)$,
encoding chemical identity prior to the incorporation of geometric information.

\subsection{$SE(3)$-equivariant message passing}

Equivariance is enforced by constructing tensor features from spherical harmonics
$Y_{\ell}^{m}(\hat{\mathbf{r}}_{ij})$ and a learnable radial basis
$R^{(t)}_{k\ell_1\ell_2}(r_{ij})$ (radial envelope Bessel basis functions~\cite{batatia2022mace}), updating features via tensor contractions,
\begin{equation}
\phi^{(t)}_{ij,k,\ell_3,m_3}
=\sum_{\ell_1 m_1,\ell_2 m_2}
C_{\ell_1 m_1,\ell_2 m_2}^{\ell_3 m_3}\,
R^{(t)}_{k\ell_1\ell_2}(r_{ij})\,
Y_{\ell_1}^{m_1}(\hat{\mathbf{r}}_{ij})\,
h^{(t)}_{j,k,\ell_2,m_2},
\end{equation}
where $C_{\cdots}^{\cdots}$ are Clebsch--Gordan coefficients.
Under a rigid rotation $g\in SE(3)$, tensor channels transform covariantly via Wigner $D$-matrices,
enforcing the Hamiltonian covariance of $\mathbf{R}\mapsto\boldsymbol{\theta}_0$:
\begin{equation}
\hat{U}(g)\hat{H}(\mathbf{R})\hat{U}^\dagger(g)=\hat{H}(g\cdot\mathbf{R}),
\end{equation}
where $\hat{U}(g)$ is the second-quantized unitary representation of the spatial rotation $g$,
under which $\hat{H}$ transforms covariantly: rotating all nuclear positions by $g$ while
simultaneously applying $\hat{U}(g)$ to the electronic degrees of freedom leaves the physics unchanged.

\subsection{Body order and many-body correlations}

To capture higher-order geometric correlations within the local receptive field, we employ a body-order
expansion.
Denoting atomic basis features by $A_i$, we form $\nu$-fold products to obtain higher-order channels,
\begin{equation}
B^{(t),\nu}_{i,\eta,k,\mathcal{L},\mathcal{M}}
=\sum_{\{l_\xi m_\xi\}}
C^{\mathcal{L}\mathcal{M}}_{\eta,\{l_\xi m_\xi\}}
\prod_{\xi=1}^{\nu} A^{(t)}_{i,k,l_\xi,m_\xi},
\end{equation}
where $C^{\mathcal{L}\mathcal{M}}_{\eta,\{l_\xi m_\xi\}}$ is a generalised Clebsch--Gordan coefficient
formed by coupling $\nu$ single-particle angular-momentum channels (cf.\ the two-body $C_{\ell_1 m_1,\ell_2 m_2}^{\ell_3 m_3}$ defined in the message-passing step above),
$\mathcal{L}$ and $\mathcal{M}$ denote the total angular momentum and its projection of the
contracted output channel (distinct from the circuit depth $L$ of Note~S4),
and $\eta$ labels distinct angular-momentum coupling paths in the body-order contraction.
Increasing $\nu$ enriches the local correlation structure accessible to the classical map, analogously
to systematic many-body expansions in ACE/MACE interatomic potentials~\cite{batatia2022mace,
wang2025acegraphene}.
We use $\nu=3$, which captures the multi-reference basin structure associated with bond breaking in
the benchmarks studied here.

\subsection{Local readout compatible with a shallow locality-preserving ansatz}

To remain compatible with the circuit causal structure (Note~S4), we apply an atomic readout head:
for each atom $i$ we extract the rotationally invariant scalar channels ($\ell=0$) from the final
layer $T$ and map them to the circuit parameters for that atom's qubits,
\begin{equation}
\boldsymbol{\theta}_{i,\mathrm{pred}}
=\pi\,\tanh\!\left(\mathrm{MLP}_{\mathrm{readout}}\!\left(\mathbf{h}^{(T)}_{i,0,0}\right)\right),
\end{equation}
where $\mathbf{h}^{(T)}_{i,0,0}\in\mathbb{R}^{K_{\mathrm{ch}}}$ stacks all $K_{\mathrm{ch}}=128$
scalar channels ($\ell=0$, $m=0$) of atom $i$ from the final message-passing layer $T$.
The readout is \emph{local}: $\boldsymbol{\theta}_i$ depends only on the atom's neighborhood within
the receptive field set by $T$ and $r_{\mathrm{cut}}$, matching the causal cone of the shallow circuit.
Restricting to $\ell=0$ channels ensures the predicted angles are invariant under global rotations.
The $\tanh$ nonlinearity confines outputs to $(-\pi,\pi)$, matching single-qubit rotation gates.

\subsection{Gauge-aware training objective}

Quantum circuit parameterizations are not unique: distinct parameter vectors can implement the same
unitary.
We train $\mathcal{P}_{\mathrm{eq}}$ with a gauge-aware objective combining a parameter anchoring
term and a state-fidelity term (see Note~S7, Eq.~\eqref{eq:objective_J}),
with $\lambda=0.1$ (Table~S3).
The first term anchors the map in the correct basin in parameter space; the second term enforces
physical equivalence and is insensitive to gauge drift along near-null Hessian directions (cf.\ Note~S2).

\section{Quantum circuit architecture and geometry-to-circuit mapping}
\label{note:circuit}

We use a constant-depth hardware-efficient ansatz (HEA) with linear nearest-neighbor connectivity.
The construction uses \emph{locality}: at fixed depth $L$, the causal cone of a local observable
remains bounded in the basin-localized, curvature-controlled regime analyzed in Note~S1.

\subsection{Ansatz definition}

For an $N$-qubit register, the variational unitary $U(\boldsymbol{\theta})$ acts on
$\ket{0}^{\otimes N}$ and consists of an initial single-qubit rotation layer followed by $L$
repeated blocks.
Writing the product with the highest-index layer on the left (standard operator-product convention),
\begin{equation}
U(\boldsymbol{\theta})
=
\left(\prod_{l=L}^{1} U_{\mathrm{layer}}(\boldsymbol{\theta}_l)\right)
\bigotimes_{i=0}^{N-1}
\left(R_y(\theta_{i,0})\,R_z(\phi_{i,0})\right),
\end{equation}
so that the initial rotation layer acts first on the state and $U_{\mathrm{layer}}(\boldsymbol{\theta}_L)$
acts last.
Each repeated block is
$U_{\mathrm{layer}}(\boldsymbol{\theta}_l)
= U_{\mathrm{ent}}\,\bigotimes_{i=0}^{N-1}\!\bigl(R_y(\theta_{i,l})\,R_z(\phi_{i,l})\bigr)$
(single-qubit rotations act first, then the entangling layer),
with $U_{\mathrm{ent}}$ the brick-wall entangling layer defined below.

\subsection{Layer structure: brick-wall entanglers}

Each layer alternates local rotations with a 1D brick-wall entangling pattern.
With qubits 0-indexed, the even and odd sublattice entangling layers are
\begin{equation}
U_{\mathrm{ent}}
=
\prod_{i\in\mathrm{even}}\mathrm{CNOT}(q_i,q_{i+1})\;
\prod_{i\in\mathrm{odd}}\mathrm{CNOT}(q_i,q_{i+1}),
\end{equation}
where ``even/odd'' refers to the parity of the qubit index $i$, and the even-sublattice pass
is applied first, as shown in the circuit schematic in Fig.~\ref{fig:circuit_diagram}.
At fixed depth $L$, this yields a bounded light cone (linear spreading at $\sim2$ qubits per layer,
one qubit on each side of a local observable),
providing a controlled setting in which gradients are governed by local basin geometry rather than
global Hilbert-space volume.

\subsection{Parameter count}

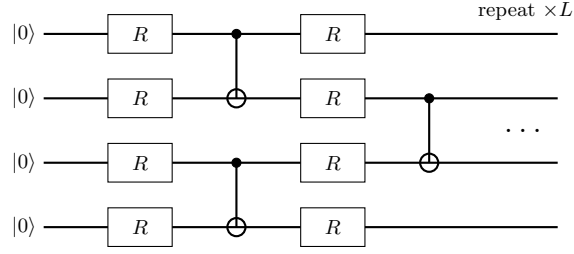
\begin{figure}[tb]
\centering
\begin{tikzpicture}[scale=0.85, every node/.style={transform shape}]
    \foreach \y in {0, -1, -2, -3} {
        \draw[thick] (0,\y) -- (8,\y);
        \node[left] at (0,\y) {$\ket{0}$};
    }
    \foreach \y in {0, -1, -2, -3} {
        \filldraw[fill=white, draw=black] (1,\y-0.3) rectangle (2,\y+0.3);
        \node at (1.5,\y) {\small $R$};
    }
    \draw[thick] (3,0) -- (3,-1);
    \filldraw[black] (3,0) circle (2pt);
    \draw[thick] (3,-1) circle (4pt);
    \draw[thick] (3,-2) -- (3,-3);
    \filldraw[black] (3,-2) circle (2pt);
    \draw[thick] (3,-3) circle (4pt);
    \foreach \y in {0, -1, -2, -3} {
        \filldraw[fill=white, draw=black] (4,\y-0.3) rectangle (5,\y+0.3);
        \node at (4.5,\y) {\small $R$};
    }
    \draw[thick] (6,-1) -- (6,-2);
    \filldraw[black] (6,-1) circle (2pt);
    \draw[thick] (6,-2) circle (4pt);
    \node at (7.5, -1.5) {\Large $\dots$};
    \node[above] at (7.5, 0.1) {\small repeat $\times L$};
\end{tikzpicture}
\caption{\textbf{Schematic of the constant-depth hardware-efficient ansatz.}
A brick-wall pattern of nearest-neighbor entanglers yields a bounded causal cone at fixed depth $L$,
providing a locality-controlled setting in which basin-localized initialization suppresses
design-like concentration effects.
Benchmarks use the shallow setting $L=4$.}
\label{fig:circuit_diagram}
\end{figure}

For $N$ qubits and depth $L$, the parameter count is
\begin{equation}
N_{\mathrm{params}} = 2N(L+1),
\end{equation}
corresponding to two rotation angles per qubit per layer (including the initial layer).
The circuit connectivity is sketched in Fig.~\ref{fig:circuit_diagram}.

\subsection{Locality-preserving interface: geometry to circuit parameters}

Let $\mathbf{v}_i$ denote the readout features associated with atom $i$ (Note~S3).
We map these features only to the rotation angles acting on qubits assigned to that atom's orbitals.
Each qubit receives two angles per layer; collecting them into a two-vector,
\begin{equation}
\begin{pmatrix}\theta_{i,l}\\\phi_{i,l}\end{pmatrix}
=\pi\,\tanh\!\left(\mathbf{v}_{\kappa}\right), \qquad \kappa=\mathrm{idx}(i,l),
\end{equation}
where $\mathbf{v}_{\kappa}\in\mathbb{R}^2$ is the two-dimensional scalar-channel readout for atom $i$
at layer $l$, the $l$-th layer segment of the full readout vector $\boldsymbol{\theta}_{i,\mathrm{pred}}$
produced by the atomic readout head of Note~S3 (\S\ref{note:mace}), with $\mathrm{idx}(i,l)$ mapping
the atom-layer pair to the corresponding channel index
$\kappa$ (distinct from the channel dimension index $k$ used in Note~S3).
This construction keeps the preconditioner within the receptive field of the shallow circuit and
uses symmetry-consistent local geometric information to set local single-qubit rotations.
The basin-attractive alignment is supplied by the supervised targets and the gauge-aware objective.

\section{Necessity of geometric equivariance: ablations and sample complexity}
\label{note:ablation}

We test whether $SE(3)$ equivariance is a structural requirement for reliable basin targeting,
rather than a heuristic convenience, using ablations on the $\mathrm{H}_4$ linear chain.
We compare $\mathcal{P}_{\mathrm{eq}}$ against an unstructured regressor that must learn rotational
structure from data, isolating the role of symmetry in reducing the effective hypothesis space.

\subsection{Learning on $SO(3)$ orbits vs.\ learning on the quotient}

An unstructured regressor operating on raw coordinates $\mathbf{R}\in\mathbb{R}^{3M}$ must infer
rotational structure from data.
For rotationally invariant targets, training budget is spent representing the
$SO(3)$ orbit of each configuration rather than the physically relevant degrees of
freedom.

\begin{figure}[tb]
\centering
\includegraphics[width=0.64\textwidth]{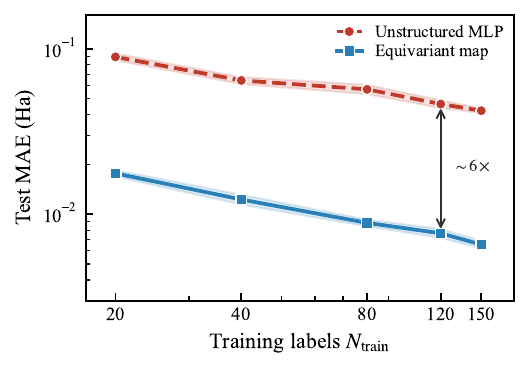}
\caption{\textbf{Equivariance reduces sample requirements on rotated $\mathrm{H}_4$.}
Generalization error on a randomly rotated $\mathrm{H}_4$ ensemble vs.\ training set size.
Unstructured model (red): slow improvement under rotation augmentation indicates sample waste on the
$SO(3)$ orbit.
Equivariant map (blue): analytic symmetry handling learns on the quotient
$\mathbb{R}^{3M}/SE(3)$ and reaches a lower mHa-scale error with fewer labels.}
\label{fig:ablation}
\end{figure}

Fig.~\ref{fig:ablation} quantifies this inefficiency.
The labels are exact FCI energies for randomly rotated and stretched $\mathrm{H}_4$ chains in
STO-3G, computed with PySCF.
Even with matched train-test splits, the unstructured model exhibits slow error decay,
indicating the burden of learning symmetry from samples.
By contrast, enforcing the state-level equivariance condition analytically, the condition stated
in the main text up to the gauge ambiguity of the circuit parameterization, makes the map
operate directly on the quotient
$\mathbb{R}^{3M}/SE(3)$.
In the data, the equivariant map reaches the mHa scale with fewer labels and lower
generalization error across the training-size scan.
The same mechanism appears in MLIP generalization: structured local
representations reduce the amount of labeled data needed to control errors outside the immediate
training set~\cite{ortner2023framework,wang2024theoretical}.

\paragraph{Physical conclusion.}
In these ablations, exact symmetry handling is required for reliable basin identification from
limited data.
Without it, the classical initialization model spends capacity on redundant symmetry directions,
leaving basin discovery in the shot-limited quantum loop.

\begin{table}[tb]
\caption{\textbf{Quantifying the symmetry gap on $\mathrm{H}_4$.}
Test MAE values (mean $\pm$ s.d.) over 10 independent train-test splits from the data used in
Fig.~\ref{fig:ablation}.}
\begin{ruledtabular}
\begin{tabular}{lccc}
Model class & Training size & Symmetry handling & Test error (mHa) \\
\hline
Unstructured (MLP) & $20$ & Raw coordinates & $89.3\pm13.9$ \\
Equivariant map & $20$ & Distance quotient & $17.6\pm2.6$ \\
Unstructured (MLP) & $150$ & Raw coordinates & $42.2\pm4.4$ \\
Equivariant map & $150$ & Distance quotient & $6.5\pm1.2$ \\
\end{tabular}
\end{ruledtabular}
\label{tab:ablation}
\end{table}

\subsection{Transfer across geometry regimes: equilibrium-trained $\to$ stretched}

We train $\mathcal{P}_{\mathrm{eq}}$ on near-equilibrium geometries ($R_{\mathrm{eq}}\pm0.2$~\AA) and
evaluate in the stretched regime ($R\gtrsim2.5$~\AA), where multi-reference character becomes dominant.
The map continues to place initialization on the correlated branch in the stretched limit
(main text PES transfer figure).
The transfer reflects a geometry-conditioned
\emph{basin correspondence}: how local bonding environments and symmetry constrain the correlated
branch, rather than merely interpolating energies within the training convex hull.
The few-label transfer protocol (pretrain on small systems, adapt readout with
$N_{\mathrm{adapt}}\approx10\text{--}50$ labeled geometries) is detailed in Note~S7.

\section{Resource accounting: shot cost model}
\label{note:cost}

The resource gain from equivariant preconditioning follows from a resource asymmetry: shot-based
energy estimation scales as $O(\epsilon^{-2})$ in the target precision $\epsilon$, whereas classical
inference incurs a fixed millisecond-scale overhead.
This note quantifies the resulting exchange rate and explains why relocating basin identification from
the quantum loop to a classical initialization map reduces the dominant cost in shot-limited
variational simulation.

\subsection{Quantum--classical resource exchange rate}

Shot noise limits the achievable statistical uncertainty of energy estimates.
To resolve an energy difference at precision $\epsilon$,
\begin{equation}
N_{\mathrm{shots}}
\sim
\frac{\mathrm{Var}(\hat{H})}{\epsilon^2},
\label{eq:S6_shots}
\end{equation}
up to constant factors determined by the measurement scheme and Hamiltonian grouping.
At chemical accuracy $\epsilon_{\mathrm{chem}}\approx1.6$~mHa, standard molecular Hamiltonian
groupings require $N_{\mathrm{shots}}\sim10^{6}\text{--}10^{7}$ shots per energy evaluation.
The dominant wall-clock cost of VQE is
\begin{equation}
C_{\mathrm{VQE}}
\propto
N_{\mathrm{shots}}\times N_{\mathrm{steps}},
\label{eq:S6_shot_tax}
\end{equation}
where $N_{\mathrm{steps}}=N_{\mathrm{steps}}^{(\mathrm{disc})}+N_{\mathrm{steps}}^{(\mathrm{local})}$
is the total iteration count, split into basin-discovery iterations
$N_{\mathrm{steps}}^{(\mathrm{disc})}$ (spent under vanishing gradients escaping plateaus)
and local-refinement iterations $N_{\mathrm{steps}}^{(\mathrm{local})}$ (within an identified basin).

A forward pass of $\mathcal{P}_{\mathrm{eq}}$ is a classical inference step with millisecond-scale
latency on commodity accelerators.
The exchange rate between one quantum evaluation at precision $\epsilon$ and one classical inference
call is $10^{4}\text{--}10^{6}$ in wall-clock time for standard hardware and
measurement strategies.
The classical preconditioning overhead is accordingly small at fixed $\epsilon$ relative to
the shot loop: it does not scale with $\epsilon^{-2}$ and remains subdominant in the time-to-solution
budget once measurement dominates.

\subsection{Wall-clock breakdown: basin discovery vs.\ local refinement}

Table~S2 gives the resource breakdown for stretched $\mathrm{N}_2$ (12 qubits), separating
(i) basin discovery (iterations spent escaping plateaus and locating a convex neighborhood) from
(ii) local refinement within an already-identified basin.
Equivariant preconditioning moves the basin-discovery phase prior to measurement, and in the
preidentified-basin protocol sets $N_{\mathrm{steps}}^{(\mathrm{disc})}=0$ while leaving
the $O(\epsilon^{-2})$ measurement scaling of the local-refinement phase unchanged.

\begin{table}[tb]
\caption{\textbf{Resource breakdown for stretched $\mathrm{N}_2$ (12 qubits).}
Representative wall-clock contributions to reach a basin-local regime ($\Delta E<10^{-2}$~Ha) at
fixed precision.
Here $N_{\mathrm{steps}}^{\mathrm{(disc)}}$ denotes iterations spent in basin discovery,
distinct from the local-refinement steps in the final row.
Equivariant preconditioning localizes initialization before the first quantum evaluation, setting
the basin-discovery cost to zero in this protocol.}
\begin{ruledtabular}
\begin{tabular}{llll}
Phase & Procedure & Scaling driver & Wall time \\
\hline
\multirow{2}{*}{Initialization} & Random draw & near zero & $\sim0$ \\
 & Equivariant preconditioning & inference latency & $\sim5\times10^{-2}$\,s \\
\multirow{2}{*}{Basin discovery} & Standard VQE (SPSA/Adam) & $N_{\mathrm{shots}}\times N_{\mathrm{steps}}^{\mathrm{(disc)}}$ &
 $\gg10^{3}$\,s$^{\dag}$ \\
 & Preconditioned VQE & $O(1)$ steps (basin pre-identified) & $\approx0$\,s \\
Local refinement & Standard / preconditioned VQE & $N_{\mathrm{shots}}(\epsilon)$ & $\sim10^{2}$\,s \\
\end{tabular}
\end{ruledtabular}
\begin{minipage}{\linewidth}
\small $^{\dag}$ Basin-discovery cost is dominated in hard instances by long plateau episodes
where gradients are exponentially small; this is the variability suppressed by basin-localized
initialization (cf.\ Fig.~\ref{fig:robustness} and Fig.~\ref{fig:S1_diagnostics}(a)).
\end{minipage}
\label{tab:cost_analysis}
\end{table}

\subsection{Amortization over a potential-energy surface scan}

For a PES scan with $K$ geometry points, standard VQE pays the basin-discovery penalty independently
at each $\mathbf{R}_i$,
\begin{equation}
C_{\mathrm{tot}}^{\mathrm{std}}
\approx
\sum_{i=1}^{K} \Bigl(N_{\mathrm{shots}}(\epsilon)\times N_{\mathrm{steps}}(\mathbf{R}_i)\Bigr),
\end{equation}
where $N_{\mathrm{steps}}(\mathbf{R}_i)=N_{\mathrm{steps}}^{(\mathrm{disc})}(\mathbf{R}_i)
+N_{\mathrm{steps}}^{(\mathrm{local})}(\mathbf{R}_i)$ exhibits heavy-tailed variability
in rugged regimes, dominated by the discovery term $N_{\mathrm{steps}}^{(\mathrm{disc})}$.
Equivariant preconditioning has a one-time training cost $C_{\mathrm{train}}$ (offline) and a
per-geometry inference cost $K\,t_{\mathrm{inf}}$,
\begin{equation}
C_{\mathrm{tot}}^{\mathrm{eq}}
\approx
C_{\mathrm{train}} + K\,t_{\mathrm{inf}}
+ \sum_{i=1}^{K} N_{\mathrm{shots}}(\epsilon)\times N_{\mathrm{steps}}^{\mathrm{(local)}}(\mathbf{R}_i),
\end{equation}
with $N_{\mathrm{steps}}^{\mathrm{(local)}}$ the small number of curvature-controlled refinement steps
once the basin is fixed.
Since the per-geometry inference latency $t_{\mathrm{inf}}\sim50$~ms is small relative to the
quantum measurement cost $N_{\mathrm{shots}}(\epsilon)\times t_{\mathrm{shot}}$ at fixed $\epsilon$,
the crossover to net advantage occurs once basin discovery constitutes a nontrivial fraction of
the quantum loop.
Training $\mathcal{P}_{\mathrm{eq}}$ amortizes basin discovery: it reduces the probability of
expensive basin-search trajectories and moves the corresponding work to a classical
preconditioner.

\section{Reproducibility parameters and numerical details}
\label{note:experiment}

This note specifies the numerical ingredients required to reproduce all reported figures and ablations,
covering (i) electronic-structure instances and reference energies, (ii) optimization protocols, and
(iii) supervised construction of $\mathcal{P}_{\mathrm{eq}}$.

\subsection{Electronic-structure instances and reference energies}

Main-text VQE benchmarks are treated in the STO-6G minimal basis.
One- and two-electron molecular integrals are generated with PySCF~\cite{sun2018pyscf},
localized with Edmiston--Ruedenberg orbitals, and assembled into the second-quantized electronic
Hamiltonian.
Statevector simulations are performed in the fixed-$(N_\alpha,N_\beta)$ determinant sector,
which gives a sparse matrix representation of the same electronic Hamiltonian used by the ansatz.
Qubit counts quoted in the main text correspond to the associated spin-orbital register; the same
second-quantized operators can be encoded by standard fermion-to-qubit transformations
such as Jordan--Wigner or Bravyi--Kitaev~\cite{seeley2012bravyi}.
The resulting electronic Hamiltonian for geometry $\mathbf{R}$ is denoted $\hat{H}(\mathbf{R})$.
The Fig.~\ref{fig:ablation} sample-complexity ablation is a separate low-dimensional regression
diagnostic: its labels are exact PySCF FCI energies for randomly rotated $\mathrm{H}_4$ geometries
in STO-3G and are not mixed with the STO-6G VQE benchmark energies.
In both cases, expensive electronic-structure evaluations supply supervised observations, as in
ML-augmented QM/MM and interatomic-model training workflows~\cite{chen2022qmml}.

\paragraph{Supervised targets.}
Circuit-parameter labels $\boldsymbol{\theta}^\star(\mathbf{R})$ are defined as parameters that
achieve the global variational minimum for $\hat{H}(\mathbf{R})$ up to parameter redundancies.
For each geometry $\mathbf{R}$ we perform basin-hopping target search followed by local refinement;
the numerical settings are listed in Table~S3.
These labels encode basin-attractive coordinates of the ground-state minimum rather than competing
local traps.

\subsection{Quantum optimization protocols}

\paragraph{Deterministic statevector regime.}
The landscape visualization in Fig.~\ref{fig:landscape} uses exact statevector simulation
and L-BFGS-B optimization~\cite{byrd1995limited}.
This regime isolates geometric obstructions of the landscape from stochastic fluctuations.
For the archived H-chain finite-gradient diagnostic, we evaluate
$N=8,12,16,20$ qubits at fixed ansatz depth $L=4$.
The $\mathrm{H}_6$ and $\mathrm{H}_8$ equivariant readout heads are adapted from the
$\mathrm{H}_4$ backbone using $14$ and $12$ labeled geometries, respectively; reported gradient
variances are averages over $15$ geometry jitters at each system size.

\paragraph{Stochastic (sampling-limited) regime.}
The disorder-robustness tests in Fig.~\ref{fig:robustness} emulate shot-noise-limited
optimization using SPSA~\cite{spall1992multivariate} with standard decaying schedules
$\alpha_k=a/(k+A)^{\alpha}$ and $c_k=c/(k+1)^{\gamma}$.
SPSA parameters are listed in Table~S3.

\subsection{Supervised construction and training protocol for $\mathcal{P}_{\mathrm{eq}}$}
\label{sec:sm_data_training}

\paragraph{Gauge-aware objective functional.}
Because circuit parameterizations contain redundant directions, we minimize a gauge-aware objective
combining parameter anchoring with physical-equivalence enforcement:
\begin{equation}
\mathcal{J}(\boldsymbol{\theta};\boldsymbol{\theta}^\star)
=
\|\boldsymbol{\theta}-\boldsymbol{\theta}^\star\|^2
+
\lambda\Bigl(1-\bigl|\langle\psi(\boldsymbol{\theta})|\psi(\boldsymbol{\theta}^\star)\rangle\bigr|^2\Bigr),
\label{eq:objective_J}
\end{equation}
with $\lambda=0.1$ unless stated otherwise.
The first term selects a fixed representative among redundant coordinates; the fidelity term
enforces physical equivalence and stabilizes training when multiple parameter representatives exist
for the same basin.

\paragraph{Geometries, splits, and few-label adaptation.}
Training geometries are drawn from the same families and bond-stretch ranges used in the main
benchmarks, with held-out geometries reserved for evaluation.
We report held-out errors rather than training losses because extrapolative reliability in
machine-learned atomistic models is tied to calibrated out-of-sample error, not only in-sample
fit quality~\cite{ho2026flexible}.
We use a two-stage procedure: (i)~pretrain $\mathcal{P}_{\mathrm{eq}}$ on small systems
($\mathrm{H}_2$, $\mathrm{H}_4$, LiH) to learn transferable local geometric features obeying
$SE(3)$ symmetry; (ii)~for a target system (e.g., stretched $\mathrm{N}_2$), freeze the lower
equivariant layers and adapt only the readout head using $N_{\mathrm{adapt}}\approx10\text{--}50$ labeled geometries.
This reduces label-generation cost while preserving the symmetry constraints responsible for basin
localization.
The second stage is the quantum-ansatz analogue of fine-tuning a pretrained atomistic foundation
model on a small task-specific labeled set~\cite{liu2026finetuning}.

\subsection{Complete reproducibility settings}

All architectural parameters and numerical settings are reported in Table~S3.

\begin{table}[tb]
\caption{\textbf{Reproducibility settings.}
Numerical settings for quantum optimization and for the equivariant preconditioner.}
\begin{ruledtabular}
\begin{tabular}{ll}
Setting & Value \\
\hline
\multicolumn{2}{l}{\textit{Quantum (VQE)}} \\
Ansatz depth (stretched $\mathrm{N}_2$) & $L=4$ (brick-wall HEA) \\
Optimizer (deterministic) & L-BFGS-B (tolerance $10^{-9}$~Ha) \\
Optimizer (stochastic) & SPSA ($a=0.1$, $c=0.1$, $A=10$, $\alpha=0.602$, $\gamma=0.101$) \\
\multicolumn{2}{l}{\textit{Supervised target generation}} \\
Target search & Basin-hopping ($T=0.5$, 100 steps, $K_{\mathrm{restart}}=10$ restarts) \\
Local refinement & BFGS (convergence $10^{-9}$~Ha) \\
\multicolumn{2}{l}{\textit{Equivariant preconditioner ($\mathcal{P}_{\mathrm{eq}}$)}} \\
Body order & $\nu=3$ (up to three-body correlations) \\
Irreps (channels) & $128\times0e$ (even-parity scalars), $128\times1o$ (odd-parity vectors) \\
Radial cutoff & $r_{\mathrm{cut}}=5.0$~\AA \\
Optimizer & AdamW~\cite{loshchilov2019decoupled} (weight decay $10^{-4}$) \\
Learning-rate schedule & Cosine annealing with warmup ($10^{-3}\to10^{-6}$) \\
Objective functional & Eq.~\eqref{eq:objective_J} with $\lambda=0.1$ \\
\end{tabular}
\end{ruledtabular}
\label{tab:repro_settings}
\end{table}

\clearpage
\makeatletter
\def\bibfont{\footnotesize}
\makeatother

\end{document}